\documentclass{optica-article}

\journal{opticajournal} 

\articletype{Research Article}

\usepackage{lineno}

\usepackage{placeins}
\usepackage{dblfloatfix} 
\newcommand{\varA}[1]{{\operatorname{#1}}}

\begin{document}

\title{Adaptive Super-Resolution Imaging Without Prior Knowledge Using a Programmable Spatial-Mode Sorter}

\author{Itay Ozer,\authormark{1,3,\textdagger, *} Michael. R. Grace\authormark{1,2,\textdagger},Pierre-Alexandre Blanche\authormark{1} and Saikat Guha\authormark{1,3}}

\address
{\authormark{1}James C. Wyant College of Optical Sciences, University of Arizona, Tucson, AZ 85721, USA\\
\authormark{2}Physical Sciences and Systems, RTX BBN, Cambridge, MA 02138, USA\\
\authormark{3}Department of Electrical and Computer Engineering, University of Maryland, College Park, MD 20742, USA\\
\authormark{'\textdagger'}Both authors contributed equally to this work}

\email{\authormark{*}iozer@umd.edu} 


\begin{abstract*} 
We consider an imaging system tasked with estimating the angular distance between two incoherently-emitting, identically bright, sub-Rayleigh-separated point sources, without any prior knowledge of the centroid or the constellation and with a fixed collected-photon budget. It was shown theoretically that splitting the optical recording time into two stages---focal-plane direct imaging to obtain a pre-estimate of the centroid, and using that estimate to center a spatial-mode sorter followed by photon detection of the sorted modes---can achieve lower mean squared error in estimating the separation~\cite{Grace:20}. In this paper, we demonstrate this in a proof-of-concept, using a programmable mode sorter we have built using multi-plane light conversion (MPLC) using a reflective spatial-light modulator (SLM) in an emulated experiment where we use a single coherent source to characterize the MPLC to electronically piece together the signature from two closely-separated quasi-monochromatic incoherent emitters. We show an improvement in estimator variance when compared to direct imaging, in good agreement with simulations.

\end{abstract*}

\section{Introduction}
 Adaptive optical imaging is an increasingly prevalent technology, enabling automated imaging systems in diverse fields such as biomedicine~\cite{barentine2023integrated}, astronomy~\cite{hampson2021adaptive}, remote sensing, and autonomous navigation. Adaptive imaging refers to the manipulation of optical hardware during an acquisition, often in a multi-stage procedure, such that processed data from initial stage(s) determine the imaging configuration used in subsequent stage(s). While dynamic scenes often necessitate adaptive imaging schemes, optical adaptation is also a powerful tool for static imaging, allowing for preliminary acquisition of "nuisance" information that is not critical to the final imaging result but helps inform the appropriate manipulations that will best perform the desired imaging task \cite{VanTrees2010,berger1999integrated,Grace:20}

For both adaptive and non-adaptive imaging, directly passing light from the scene through conventional optics like lenses and mirrors before reaching a focal-plane detector array constitutes the most widespread method. While effective for many purposes, this "direct imaging" approach encounters a resolution constraint often referred to as the Rayleigh limit \cite{Rayleigh}, an angular distance scale beyond which the features of the scene become exceedingly hard to resolve. The limitation in resolution stems from diffraction effects arising due to the finite aperture of the imaging system, as illustrated in Fig.\ref{fig:Experiment_reason}. For a telescope-like system with a circular aperture, the Rayleigh limit sets a characteristic length scale
\begin{equation}
   \theta_{\rm{res}}=1.22\times\frac{\lambda\times f}{D},
   \label{eq:theta_res}
\end{equation}
where $\lambda$ is the wavelength of the incoming light, $f$ is the focal length of the optical system, and $D$ is the diameter of the entrance pupil of the optical system, which is dictated by the finite circular aperture. An analogous diffraction limit is encountered in microscopy due to the finite numerical aperture of a circular objective lens.

While sub-diffraction imaging can be done in the near field \cite{harootunian1986super}, with controllable fluorescent probes \cite{hell1994breaking,betzig2006imaging,rust2006sub,hess2006ultra}, and via structrured optical excitation \cite{gustafsson2000surpassing}, fully passive optical super-resolution was made possible by a recent class of imaging receivers known as spatial mode demultiplexing (SPADE) \cite{19}. SPADE uses physical linear optical processes such as multi-path interferometry \cite{Nair:16} or holography \cite{Paur:16} to couple specific orthogonal spatial modes of the incoming optical field to corresponding intensity detectors. References present various experimental demonstrations showcasing resolution enhancement below the Rayleigh limit in the transverse distance estimation problem through the utilization of spatial mode demultiplexing to classify Hermite-Gauss (HG) modes prior to detection {\cite{Tan:23,Rouviere:24,Santamaria:23,Boucher:20}}. Recently, multi-plane light conversion (MPLC)~{\cite{Fontaine_Ryf_Chen_Neilson_Kim_Carpenter_2019,morizur2010programmable,labroille2014efficient}}, which uses a succession of phase masks to transform the optical wavefront {(Fig.~\ref{fig:SRONN Set-up}(a))}, has emerged as {arguably} the most promising SPADE implementation{\cite{Tan:23,Rouviere:24,Santamaria:23,Boucher:20}}. Notably, a commercially available MPLC device that demultiplexes a fixed set of 2D spatial modes (Hermite Gauss modes) has spurred experimental progress for several recently demonstrated sub-diffraction imaging tasks \cite{Boucher:20,Santamaria:23,Grenapin23,Rouviere:24}. 

Neither SPADE nor MPLC are constrained to a particular set of spatial modes, and a device that can update operational spatial mode basis would be a perfect fit for adaptive super-resolution imaging. Several theoretical works have proposed using mid-acquisition adaptation to extend passive super-resolution techniques to more complex imaging tasks~\cite{Grace:20,de2021discrimination,bao2021quantum,Kwan23}. Experimental progress in adaptive mode sorting is limited to date, with some systems demonstrating limited adaptability~\cite{fickler2017custom}. A fully adaptive MPLC implementation could perform highly complex imaging tasks in the presence of nuisance parameters{\cite{Kwan23,Matlin2022}}. 

Our experiment attempts to complete a more complicated task, demonstrating resolution enhancement below the Rayleigh limit in the transverse distance estimation problem without prior knowledge of the locations of the two {identically bright,} incoherent sources or the centroid of the scene \cite{Grace:20}. {In addition, we employ a spatial light modulator (SLM) to implement a reconfigurable mode sorter, enabling an adaptive two-stage measurement procedure that uses the results of a first-stage direct imaging measurement to algorithmically align the second-stage SPADE device. Using an adaptive optical device (e.g., an SLM) to replace an optomechancial pointing apparatus has significant potential advantages for real-world imaging systems including size, weight and power (SWaP) considerations~\cite{martin2018photonic}, alignment latency (60 Hz refresh rate for the SLM in our experiment), and pointing stability, especially considering the rapid development of compact, tunable metasurfaces for optical modulation~\cite{fang2024nonvolatile}.} 


\section{Main Results}
\begin{figure}[ht]
\centering
\fbox{\includegraphics[width=.75\linewidth]{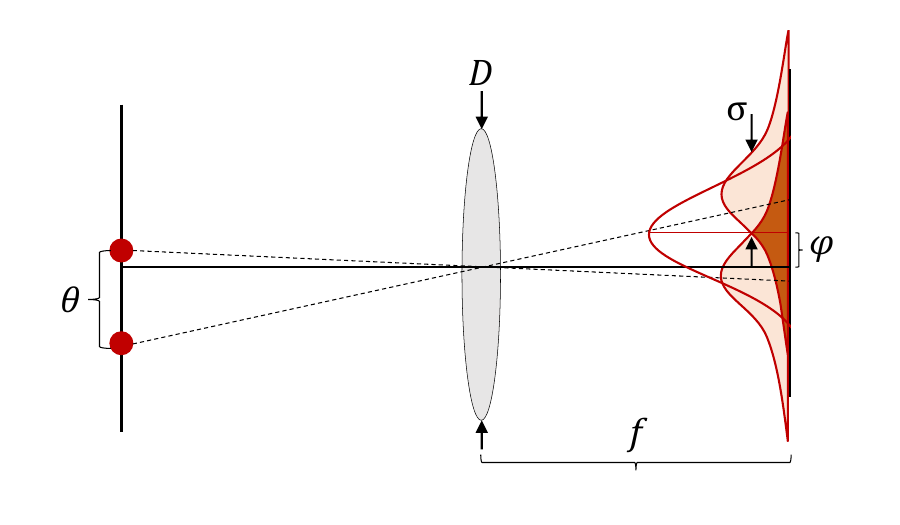}}
\caption{Example image plane intensity distribution from two point sources with separation $\theta$, centroid $\phi$, aperture diameter $D$, focal length $f$, and PSF width $\sigma$.}
\label{fig:Experiment_reason}
\end{figure}
Our experiment aims to estimate the separation ($\theta$) between two identical incoherent point sources with an unknown centroid ($\phi$), illustrated in Figure \ref{fig:Experiment_reason}. To accomplish this, we adopted the strategy that was first hinted in \cite{19} and fully outlined in \cite{Grace:20}, which suggests dividing photon resources between direct detection for centroid estimation and binary mode sorting for separation estimation. Direct detection has the advantage of shift invariance, allowing it to identify the location of the source pair over a wide field of view, and also is close to an optimal measurement of the object centroid \cite{Aqil}. Our second-stage SPADE measurement has been shown to achieve the resolution quantum limit for estimating small separations between point sources when the mode sorter is nearly perfectly aligned \cite{19}. {Theoretically, the optimal ratio for dividing the photon resources is separation ($\theta$) dependent and can be determined adaptively, which was demonstrated via simulations in \cite{Grace:20}. In our experiment, we divided the photons equally between the two stages for simplicity, while adaptively utilizing the centroid estimate from first-stage direct imaging to align the second-stage mode sorter. }

The principal findings, expounded upon in the subsequent results section, reveal notable enhancements. Specifically, our hybrid approach demonstrates a reduction in variance of separation estimation for separations of up to {a factor of 0.3 of the Rayleigh criterion}, as well as a decrease in mean squared error of separation estimation for separations of up to textcolor{red}{a factor of 0.2 of the Rayleigh criterion} when contrasted with direct imaging. The estimation of separations is executed using maximum likelihood estimators applied to both the direct imaging data and the sorted data.

It is important to note that the experimental results are partly emulated, both in that we did not employ two distinct incoherent sources simultaneously but computationally combined data that were acquired sequentially from individual point sources{, and also due to the fact that we did not implement real-time adaptation but performed the two stages of our measurement asynchronously. Nevertheless, the findings outlined in this paper serve as a proof of concept confirming the experimental viability of adaptive spatial mode sorting for sub-diffractive imaging.}


\section{System Overview}
\begin{figure}[h!]
\centering
\fbox{\includegraphics[width=\linewidth]{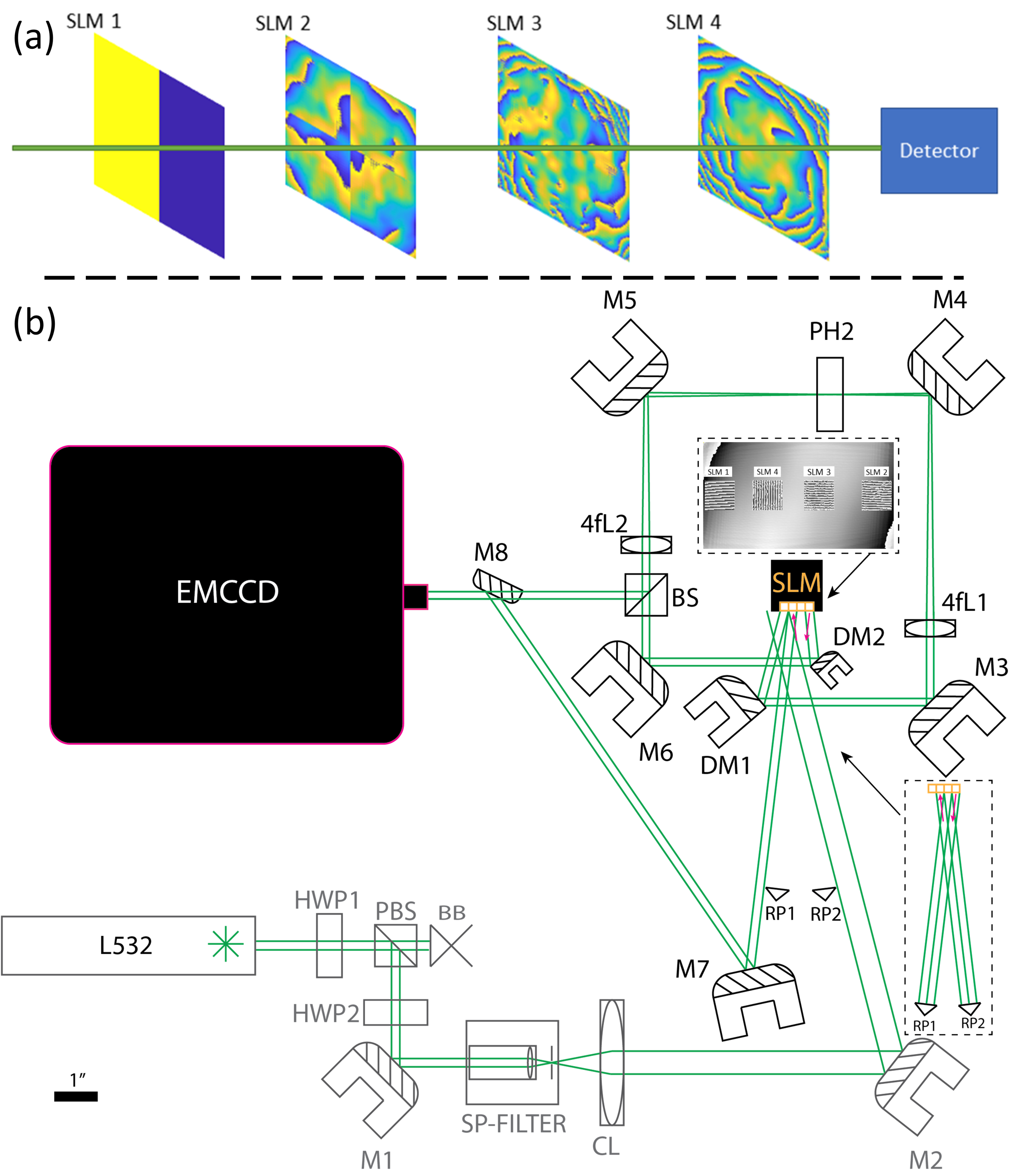}}
\caption{(a) Sample phase masks that would be displayed on the SLM, where SLM 1 acts as the object under test and the following three masks act as the MPLC masks. (b) Experimental setup. L532: 532 nm laser. HWP: half wave plate. PBS: polarized beam splitter. BB: beam block. M: mirror. SP-FILTER: spatial filter for single-mode beam processing. CL: cylindrical lens. SLM: spatial light modulator. DM: D-shaped pickoff mirror. 4fL: relay lens for 4f system. PH: pinhole. BS: non-polarzing beamsplitter. RP: right-angle prism. M8: flip mirror. EMCCD: electron-multiplying charge-coupled device camera.}
\label{fig:SRONN Set-up}
\end{figure}
We constructed a system that is able to acquire direct imaging data from the scene under test, estimate the scene's centroid, align itself to the found centroid, and sort the appropriate orthogonal mode basis. \\
It is important to note that this setup is capable of more than separation estimation between 2 point sources. However, in this paper, we will focus on this specific application.
\subsection{System walkthrough}
Our experimental setup consists of 4 main subsystems:\\
\textbf{Illumination:} Our 532 nm laser source (Edmund Optics 10 mW DPSS) first goes through a linear polarizer in order to align its polarization to the polarization required by the SLM. A spatial filter is then used to eliminate the higher spatial frequencies of the beam and produce a clean, expanded, and uniform, Gaussian beam. The filtered {and expanded} beam is then collimated and directed to the SLM. {Our goal is to use the expanded gaussian beam to uniformly illuminate the first SLM region (trying to get as close to a flat top). This will yield a scene with uniform intensity distribution. 
} \\
\textbf{SLM:} In our system we use a liquid crystal based SLM (Holoeye PLUTO-2.1-VIS-016{, 8 $\mu$m pixel pitch), which applies a polarization-sensitive phase modulation to incident light at 532 nm in response to pixel-dependent, digitally driven voltage inputs.} As can be seen in Fig.\ref{fig:SRONN Set-up}(a), four square regions {(300x300 pixels)} are displayed on the SLM. The first region is used to encode an {intensity}-modulated object, and the other three are used as phase masks created by our algorithm, which was based on \cite{Fontaine_Ryf_Chen_Neilson_Kim_Carpenter_2019}, forming the MPLC-based optical transformation that will ideally direct each orthogonal input mode to a respective, pre-set, location on the detector. On each square, we applied a blazed grating {(period of 4.2 pixels on SLM region 1, period of 8 pixels on SLM regions 2-4)} and {coupled the first diffraction order of each blazed grating into the beampath} rather than the zeroth order. The zeroth order is noisier due to the reflection of unmodulated light, while the first diffraction order is lossier. We use the latter as our focus is to minimize the crosstalk in mode sorting. {The blazed grating on the first SLM region was used to generate binary-encoded intensity modulated objects by only coupling light into the continuing beampath from the portions of the SLM region containing the blazed grating. The blazed grating on the second, third and fourth SLM regions were statically displayed across the whole region over top of the MPLC phase masks. We additionally applied a correction phase mask over top of the entire SLM display to correct the systematic static wavefront aberration we characterized for our SLM device (see Supplementary Material Section 2)}. Note that due to the SLM's programmability and the fact that our MPLC algorithm works for any orthogonal mode set, we achieve a reconfigurable mode sorter {(see Supplementary Material Section 1 for sorting performance using different orthogonal mode sets)}.   \\
\textbf{4f system:} The light next goes through a 4f system {consisting of two plano-convex lenses of focal length $f=200$ mm} right after the object is encoded at the first bounce from the SLM. The purpose of the 4f system is to introduce diffractive blur on the {intensity}-encoded object by placing a pinhole (PH2 in Fig.\ref{fig:SRONN Set-up}(b)) at the Fourier plane of the 4f system to effectively create a low pass {spatial} filter. By choosing the diameter of PH2 (in our experiment we used a pinhole with diameter {$D=200 \mu$m}), we can change the width of the PSF. {This allows us to use our tabletop optical setup to conduct a super-resolution imaging experiment with a characteristic PSF width of $\theta_{\rm res}=1.22*532\textrm{ nm}*200\textrm{ mm}/200\, \mu \textrm{m}=649\, \mu \textrm{m}$, equal to 81.125 SLM pixels or 40.56 camera pixels. The conjugate planes at the start and the end of the 4f system are coupled to the first and second SLM regions, respectively, allowing the intensity-encoded object to be presented to the remaining SLM planes for MPLC processing.}\\
\textbf{MPLC:} {Our 3-plane MPLC uses successive reflections off of the last three SLM regions. Two right angle prisms (RP) placed at a distance of 100 mm redirect the light back to the next SLM region. Prisms were used instead of a single mirror reflector in order to achieve an optical design that separately accesses SLM region 1 for object modulation. The MPLC algorithm from \cite{Fontaine_Ryf_Chen_Neilson_Kim_Carpenter_2019} been modified to account for this geometry.}\\
\textbf{Detection:} Finally, we use a beam splitter (BS) to be able to access both a direct image and the mode demultiplexed detection data. Note that only one of the outputs is going to reach the camera at a time and this will be decided by the flip mirror (M8 in Fig.\ref{fig:SRONN Set-up}(b)). {The camera is positioned on the table such that the detector plane lies at a conjugate image plane to the the first SLM region (with unit magnification) when M8 is flipped down and at the programmed output plane of the MPLC when M8 is flipped up.} Each of these outputs is detected using the photon counting feature of an electron-multiplying charge-coupled device (EMCCD) camera (Andor iXon 897{, 16 $\mu$m pixel pitch}). {See Section 4 of the Supplementary Material for details on photon counting characterization.}

A more detailed system walk-through of our experimental setup can be found in \cite{Ozer:thesis}.

\subsection{Zernike Modes Sorting Characterization}
{Since we used a circular pinhole as our aperture, our imaging system has a 2D Airy disk PSF. The associated coherent PSF has the functional form
\begin{equation}
    \psi(x,y)=\frac{J_1\left(1.22\pi r/\theta_{\rm res}\right)}{0.61\pi r/\theta_{\rm res}}
    \label{eq:AiryPSF}
\end{equation}
where $J_1(\cdot)$ is the Bessel function of the first kind and where
\begin{equation}
    r=\sqrt{x^2+y^2}.
    \label{eq:r}
\end{equation}
The characteristic resolution scale $\theta_{\rm res}$ [Eq.~\eqref{eq:theta_res}] is defined to match the radial distance to the first zero in the Airy ring diffraction pattern.} 

{For estimating the separation between two point sources using a circularly symmetric imaging aperture, an optimal modal basis for SPADE processing is the PSF-adapted (PAD) basis~\cite{Kerviche2017a,Rehacek2017b}. This basis is constructed by first taking spatial derivatives of the coherent PSF along the $x$ and $y$ dimensions and subsequently performing a Gram-Schmidt orthogonalization procedure to arrive at an orthonormal basis. For our 1D imaging experiment, we only considered spatial derivatives along one dimension in our modal distribution and term the $m^{\rm th}$ mode "PAD0$m$." We note that we did not choose to sort Zernike polynomials, which are another valid choice of orthonormal basis for quantum-optimal estimation in this scenario~\cite{YuPrasad2019}, because they do not factor into $x$ and $y$ components and are therefore inconvenient for 1D parameter estimation.} 

\begin{figure}[ht]
\centering
\fbox{\includegraphics[width=\linewidth]{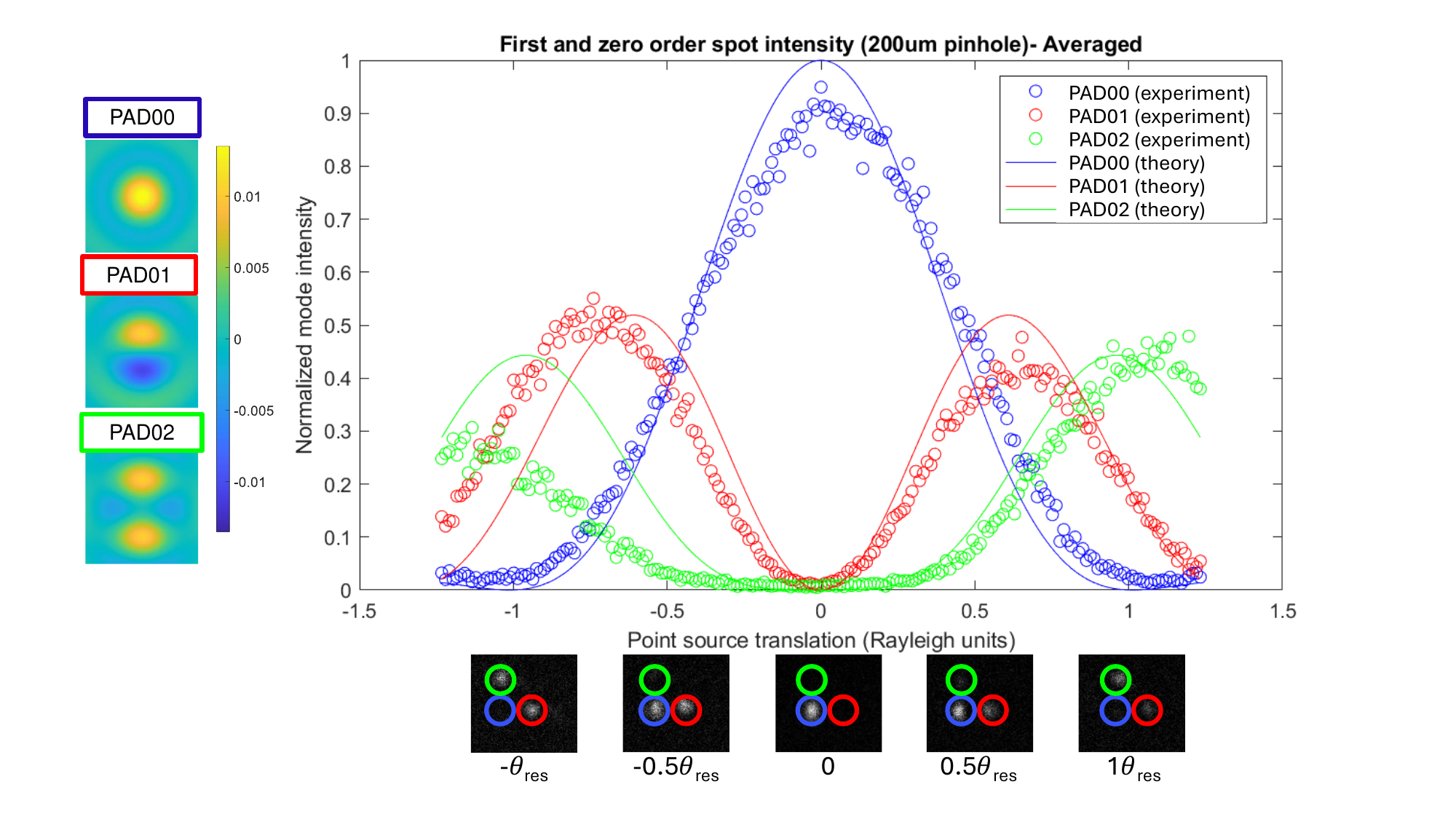}}
\caption{Calibration plot of theoretical (solid lines) and experimental (circles) intensities found in each mode with respect to the distance of the point source away from the known center {The step size was discretized by the SLM pixels (each pixel equate to $\approx$ 0.0125 Rayleigh units) }. Amplitudes of the three orthogonal spatial modes that are sorted by the MPLC are shown on the left. Below the plot are the raw experimental images with the regions designated for each mode shown in circles, where each image corresponds to the indicated distance away from the system's axis. Note: we report crosstalk value of {1.69\%} between the fundamental mode and the higher order modes when the point source translation=0.}
\label{fig:Lookup_table}
\end{figure}

To ensure the feasibility of the experiment we tested how well our mode sorter sorts the {PAD} modes of interest. To conduct that test we displayed a {blazed grating on only a 20x20 pixel square on the first SLM region (SLM 1 in Fig.~\ref{fig:SRONN Set-up}a) to approximate a point source} and swept it across the whole scene. {We chose a 20x20 pixel size for our point source because it was empirically the smallest region we could designate for a blazed grating before a sharp dropoff in the reflectivity into the first blazed grating order. In Fig.~\ref{fig:Lookup_table} we plot the experimentally characterized mode intensity distribution as a function of point source spatial translation, where the term "Rayleigh units" means that the x-axis is scaled by the distance to the first zero in the Airy disk. The experimental data from the point source sweep closely matches the theoretical expectation, which was numerically calculated from the overlap of the shifted Airy disk with especially when the deviation off-axis is small. When the object is exactly on axis (where theoretically we expect all the photons to be in the PAD00 mode) we saw only 1.69\% cross-talk (definition of cross-talk further discussed in the Supplementary Material Section 1).}

Our experiment sought to imitate the binary mode sorting case discussed in~\cite{Grace:20}. However, as the separation between the point sources increased we saw more contribution from higher order modes (as expected) {and any light from unsorted modes will randomly spread across the camera after passing through the MPLC, contributing noise to the measurement}. To mitigate this contribution we decided to sort 3 modes {(PAD00, PAD01, and PAD02, see left side of Fig.~\ref{fig:Lookup_table})}. The objective of sorting the third mode {(PAD02)} was to ensure that the photons that live in the higher-order modes (especially when the separation increases) are directed there rather than to the other two spots which would corrupt the data {(see bottom of Fig.~\ref{fig:Lookup_table})}. The other two spots {(PAD00 and PAD01)} were then used to evaluate the binary mode sorter's performance in super-resolution imaging. {The decision to sort only 3 modes is based on the fact that sorting more modes hurts the performance of the mode sorter and produces more crosstalk (see Supplementary Material, Section 1).}


\section{The Experiment}
\subsection{Experiment Overview}
The objective of our experiment is to prove the capability of our mode sorting system to achieve super-resolution parameter estimation with relaxed prior information. {More specifically, we sought to demonstrate the theoretical conclusion of Ref.~\cite{Grace:20}: that adaptive spatial mode sorting can exceed the two-point-source separation estimation precision of direct imaging without knowing the object centroid \emph{a priori}. We conducted our 2-stage experiment using the following procedure, of which we performed 200 independent trials for each of twelve different point source separations ($\theta$) ranging from 0.1 Rayleigh units to 1.2 Rayleigh units, resulting in 2400 trials each for our two-stage adaptive receiver and for direct imaging alone. For each one of these 4800 trials, the centroid ($\phi$) was randomly sampled from a uniform distribution across a domain of 100 SLM pixels (1.23 Rayleigh units). The two point sources for a given trial were therefore located at $\phi+\theta/2$ and $\phi-\theta/2$, which correspond to 1D pixel values on the SLM.}

First, direct imaging data was acquired and used to estimate the centroid of the two point sources (Section 5A).  We then use the centroid estimate as an input to our MPLC algorithm~\cite{Fontaine_Ryf_Chen_Neilson_Kim_Carpenter_2019} to generate phase masks that sort PAD modes that are centered at the location of the centroid estimate in the object plane. This adaptive generation of a special-purpose set of phase masks for each trial effectively aligns our system to the scene and eliminates any need for physical manipulations.  We then used the MPLC formed by the computed phase masks to perform the second-stage SPADE measurement. From this acquired data, we estimate the separation between the two point sources using maximum likelihood estimation as part of the data processing.

All the data in this experiment was taken using the photon counting mode in our EMCCD and the total number of photon arrivals for each trial was aimed to be 10,000. {This means that each of the two stages detected 5,000 photons, on average, whereas the direct imaging measurement used for comparison detected all 10,000 photons.} The laser power going into the system was calibrated accordingly while ensuring a low likelihood of multiple photons events. (See supplementary materials section {4}). We conducted the experiment using a laser (coherent source), while for the experiment to correctly correlate the theory of super-resolution for two incoherent point sources, the light from the two point sources cannot interfere with each other. We therefore acquired data from only one point source at a time and then combined the two images computationally.  

\subsection{First Stage}
 \begin{figure}[htbp]
\centering
\fbox{\includegraphics[width=10cm]{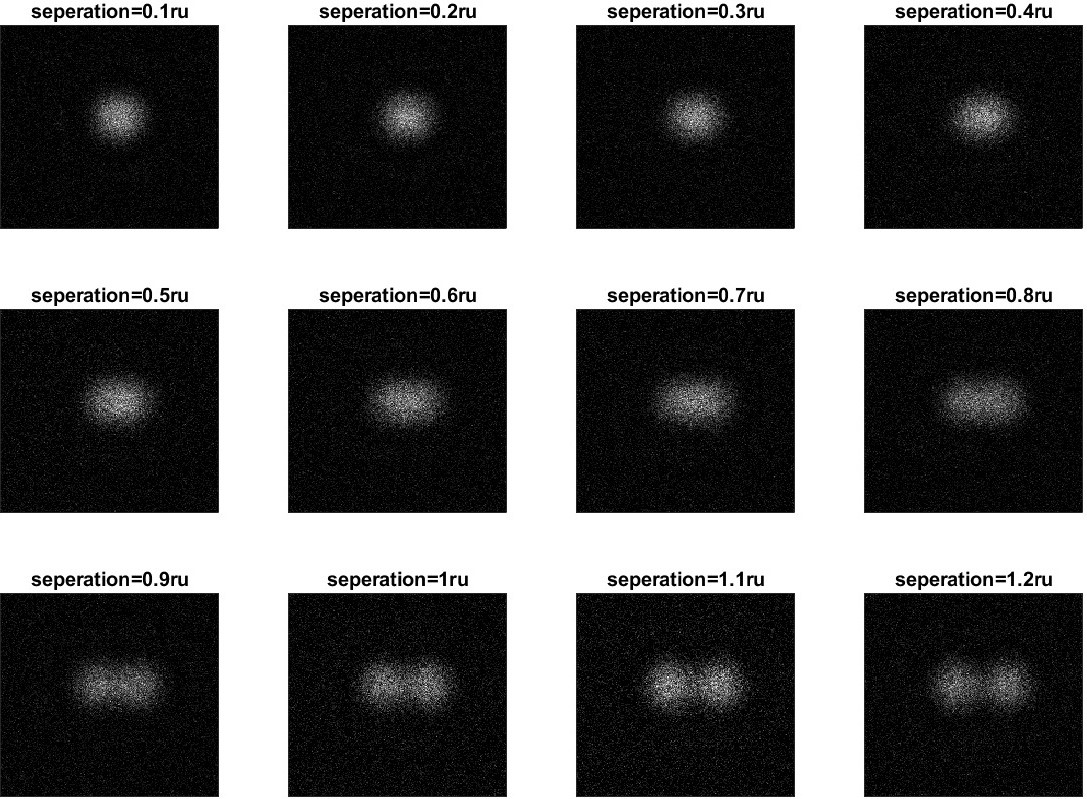}}
\caption{Example output from direct imaging with different separations. As the separation increases it is easier to notice that two point sources are present.  }
\label{fig:direct_output}
\end{figure}
In the first stage of this experiment, we acquired direct imaging data by sweeping the point source (20x20 pixels on the SLM, {limited by the amount of light needed in the system}) across the full field of view (200 pixels vertically) 5 times. {At this point it is important to recall that this was an emulated experiment, therefore, 5 sweeps allowed us to combine images from different sweeps to create our emulated, 2 point source scene. Combining images from different sweeps to construct our scene will make it more difficult to estimate both the centroid and the separation and will make this emulated scenario more realistic.} Once all the data was acquired, we randomly chose 200 different centroids and for each random centroid we chose the appropriate frame from a random data set (each one of the 5 point-source sweeps is considered one data set). After the chosen frames were combined into one image (an example of that with all tested separations is shown in Fig.\ref{fig:direct_output}), we estimated the centroid using the center of mass method.

After the centroid was estimated, it was fed into to the MPLC phase mask generation algorithm which created the appropriate phase masks to sort the wanted modes, where the estimated centroid is set to be the center of the scene for each computation. A total of 2400 (12 separations and 200 data points for each separation) centroids and phase mask sets were generated. {While the computation time required to generate a set of phase masks for each trial necessitated that we operate our two-stage experiment asynchronously, in a future system a set of phase masks could be chosen from a pre-computed set based on the estimate from the preliminary measurement.}

 \subsection{Second Stage}
\begin{figure}[htbp]
\centering
\fbox{\includegraphics[width=10cm]{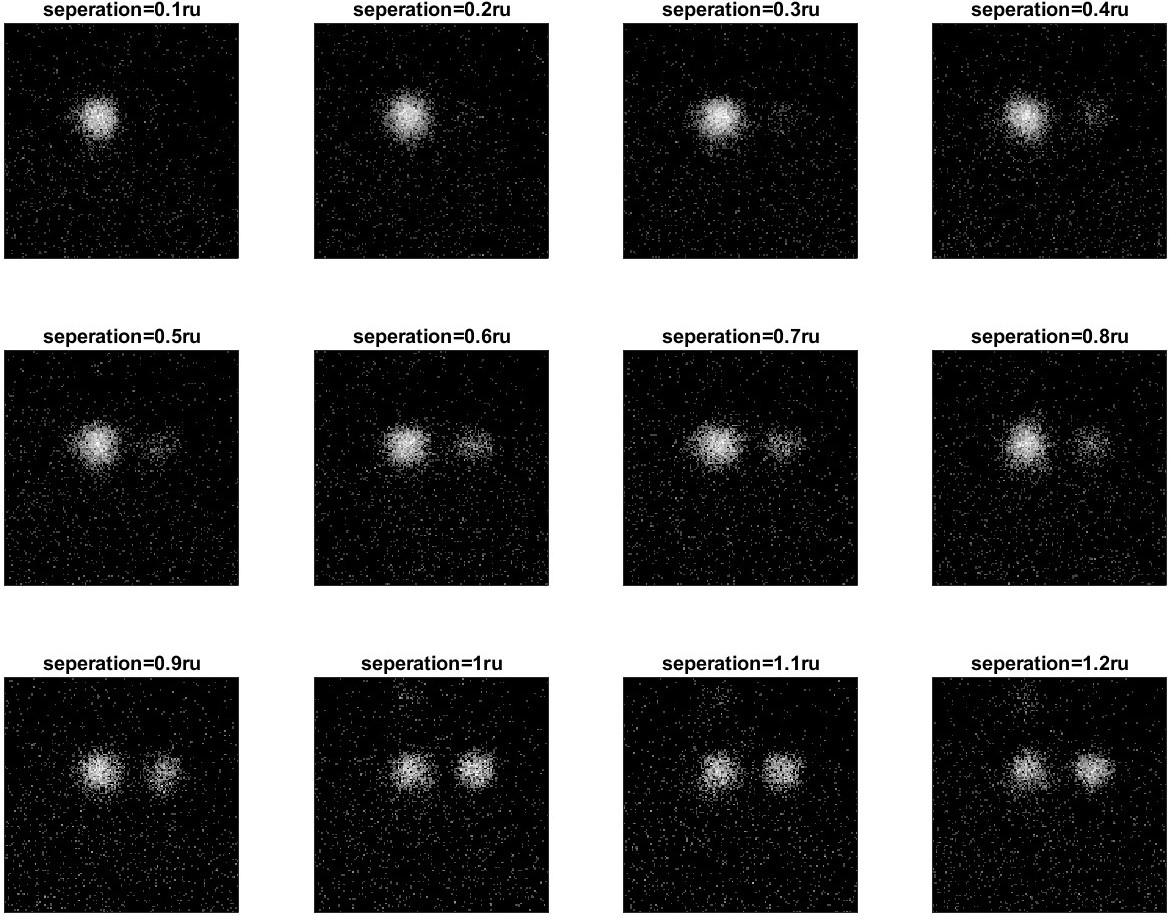}}
\caption{Example output from sorting. The left spot represents the relative presence of the 0th order mode, the right spot represents the relative presence of the 1st order mode, and the upper left spot (dim and only present in higher separations) represents the presence of the 2nd order mode.}
\label{fig:sorting_output}
\end{figure}
 The second stage of the experiment focused on acquiring the mode sorting data (shown in Fig.\ref{fig:sorting_output}). In this stage, we {computationally combined the images acquired from displaying the two point sources separately, where the data was the camera acquisitions with the MPLC active.} This was done 4800 times (each data point required 2 point sources, and we did it 200 times for the 12 separations that were tested).

 \section{Data Processing}
 

 \subsection{Centroid Estimation}
 To estimate the centroid we used the direct imaging data (as shown in Fig.\ref{fig:direct_output}) from different trials, as mentioned in the \emph{First Stage} sub-section. Once we had the computationally combined image we used the \emph{center of mass} method to estimate the centroid. 

 Our detector had 512x512 pixels so the center of the camera was set to be the zero position. Each pixel  to the left for the x centroid ($\phi_x$), or down for the y centroid ($\phi_y$) was assigned a value from -256-0 accordingly. Similarly, each pixel to the right for $\phi_x$ or up for $\phi_y$ was assigned a value from 0-256 accordingly. This gave us two vectors:\\ $V_x,V_y=[-255:256]\times Pixel Pitch$. Using these vectors in a mesh grid ($V_x \rightarrow M_{x_{i,j}}$, $V_y \rightarrow M_{y_{i,j}}$) along with the acquired data ($Im_{i,j}$) we can calculate the estimate of the x and y centroids as such:
 \begin{subequations}
\begin{align}
    \label{eq:subeq1a}
    \hat{\phi}_x=\frac{\sum_{i=-256}^{i=255}\sum_{j=-256}^{j=255}  M_{x_{i,j}}\times Im_{i,j}}{\sum_{i=-256}^{i=255}\sum_{j=-256}^{j=255}Im_{i,j} }, \rm{and} \\
    \label{eq:subeq1b}
    \hat{\phi}_y=\frac{\sum_{i=-256}^{i=255}\sum_{j=-256}^{j=255}  M_{y_{i,j}}\times Im_{i,j}}{\sum_{i=-256}^{i=255}\sum_{j=-256}^{j=255}Im_{i,j} }.
\end{align}
\end{subequations}

 \subsection{Separation Estimation- Direct Imaging}
Because in this experiment we have two point sources separated in 1D, we can make use of our $x$ and $y$ centroid estimates and describe the intensity pattern for direct imaging data on the detector (conditioned on a point source separation of $\theta$) as
  \begin{equation}
     {\Psi(x,y\vert\theta)=\left\vert\psi\left(x-\hat{\phi}_x+\frac{\theta}{2},y-\hat{\phi}_y\right)\right\vert^2+\left\vert\psi\left(x-\hat{\phi}_x-\frac{\theta}{2},y-\hat{\phi}_y\right)\right\vert^2.}
 \end{equation}\\
 Using the intensity pattern on the detector as well as the estimated centroid we can write a likelihood function:
 \begin{equation}
     P_D(\{x_i\}_{i=1}^{n_1}|\theta)=\prod_{i=1}^{n_1}\Psi(x_i,y\vert\theta) ,
 \end{equation}\\
 where $x_i$ stands for the arrival position of the $i^\text{th}$ detected photon, and $n_1$ stands for the total number of photons detected during the direct imaging measurement. 
 
 {For the case when the entire integration time is used for direct imaging, the arrival locations of our photon counting procedure can be used to estimate the parameter of interest.} By maximizing the log-likelihood function we get our estimation for the point-source separation:
\begin{equation}
    \hat{\theta}_{\rm{direct}} =\mathrm{argmax}_{\theta}\left(\log[P_D(\{x_i\}_{i=1}^{n_1}|\theta)]\right) .
 \end{equation}\\

 \subsection{Separation Estimation- Sorting}
 In this paper, we present two ways in which we processed our sorting data: 1) Maximum likelihood- This is the way that we used to characterize the theoretical performance of the system as this method should give us the most accurate estimation of the parameter of interest, $\theta$ when the distribution of the data follows the model exactly.
 2) Comparison to characterization- Since our system may have non-idealities, such as slight misalignments, there is a chance that the distribution will not exactly follow the theoretical model. Therefore, in this method, we compared our raw data to a performance characterization of our system, as seen in Fig. \ref{fig:Lookup_table}. 
 \subsubsection{Maximum Likelihood}
In our experiment, we sorted PAD basis modes for the Airy disk PSF. However, for simplicity, and due to the similarity of their plots in the super-resolution regime we matched our binary SPADE photon counting results to the probability of a photon landing in the 0th and 1st Hermite-Gauss modes as a function of the separation, $\theta$, which can be described as such:

\begin{subequations}
\begin{align}
    \label{eq:subeq7a}
    \psi_0 &= {e^{-\theta^2/\theta_{\mathrm{res}}^2}}, \quad \mathrm{and} \\
    \label{eq:subeq7b}
    \psi_1 &= {2\frac{\theta^2}{\theta_{\mathrm{res}}^2} e^{-\theta^2/\theta_{\mathrm{res}}^2}} \approx 1-\psi_0,
\end{align}
\end{subequations}
The $\psi_1$ function can be approximated as it is in Eq.~\ref{eq:subeq7b} in order to model a binary SPADE, such that all the photons that will not be counted in the $\psi_0$ spot, will be counted from the $\psi_1$ spot.


 Using the description of photon counts expectation as a function of $\theta$ we can describe the sorting likelihood function $P_B$ as a binomial distribution:
 \begin{equation}
     P_B(I_0,n_2|\theta) = \begin{pmatrix}
    n_2 \\
    I_0
\end{pmatrix} \psi_0^{I_0} \psi_1^{n_2-I_0},
 \end{equation}

where $I_0$ is the number of photons detected in the $\psi_0$ spot, and $n_2$ stands for the total photons detected in the sorting stage.

Now, to estimate the most likely $\theta$ we maximize the log-likelihood function $P_B$:
\begin{equation}
    \hat{\theta}_{\rm{Sorting}}=\mathrm{argmax}_{\theta}\left(\log[P_B(I_0,n_2|\theta)]\right).
\end{equation}

\subsubsection{Comparison to Characterization}
{Another separation estimation method, we used is a direct comparison between the photon numbers detected in each mode and the expected relative intensity shown in \ref{fig:Lookup_table}.}
The exact number of photons that gets detected at every data point will vary. Because of that, rather than comparing the number of photons in each mode to our characterization we found the ratio between $\psi_0$ and $\psi_1$ from our data and matched it to the same ratio from our interpolated system's characterization (Fig. \ref{fig:Lookup_table}). By matching the ratios we could then find what is the separation that is the most fitting for the data. {Note that this method will give us a discrete, rather than a continuous set of solutions. On the other hand, when using this method, experimental deviations, using our system, away from the theoretical values don't hurt the estimation.}
 \subsection{Separation Estimation- 2-Stage}
 As mentioned in section 3A, when we try to estimate $\theta$ using our mode sorter, we still use direct imaging for about half of the photons for the centroid estimation. In this method {we use the 5000 photons in direct imaging for two purposes,} instead of just using these photons for centroid estimation we also use them for the separation estimation. This leads to a new likelihood function that utilizes all the photons collected:
 \begin{equation}
     P_{\varA{2-stage}}(\{x_i\}_{i=1}^{n_1},I_0,n_2|\theta)=P_{D}(\{x_i\}_{i=1}^{n_1},\phi_x|\theta)\times P_B(I_0,n_2|\theta).
 \end{equation}
Therefore, the estimated $\theta$ using this method can be found by maximizing the log-likelihood function for the composite data:
\begin{equation}
    \hat{\theta}_{\varA{2-stage}}=\mathrm{argmax}_{\theta}\left( \log[P_{\varA{2-stage}}(\{x_i\}_{i=1}^{n_1},I_0,n_2|\theta)]\right).
\end{equation}

{\subsection{Statistical performance evaluation}
\label{sec:Statistics}
We computed two statistical quantities to summarize the performance of our various receivers. The first is the mean squared error (MSE) of the estimation procedure with respect to the true value of the parameter $\theta$, defined for a given measurement as
\begin{equation}
    \textrm{MSE}(\hat{\theta}_{\rm meas})=\left\langle \left(\frac{\hat{\theta}_{\rm meas}-\theta}{\theta_{\rm res}}\right)^2\right\rangle.
\end{equation}
The second measure of performance is the variance of the separation estimate, defined by
\begin{equation}
    \textrm{Var}(\hat{\theta}_{\rm meas})=\left\langle \left(\frac{\hat{\theta}_{\rm meas}}{\theta_{\rm res}}\right)^2\right\rangle-\left\langle\frac{\hat{\theta}_{\rm meas}}{\theta_{\rm res}}\right\rangle^2.
\end{equation}
Another relevant summary statistic of the experimental data processing will be the estimator bias, given by (see Supplementary Materials)
\begin{equation}
    \textrm{B}(\hat{\theta}_{\rm meas})=\left\langle \frac{\hat{\theta}_{\rm meas}-\theta}{\theta_{\rm res}}\right\rangle.
\end{equation}
Note that the source separation is written in Rayleigh units for all three of these quantities. From these definitions,
\begin{equation}
    \textrm{MSE}(\hat{\theta}_{\rm meas})=\textrm{Var}(\hat{\theta}_{\rm meas})+\textrm{B}(\hat{\theta}_{\rm meas})^2.
    \label{eq:MSE_Var_Bias}
\end{equation}}

\section{Results}
 \begin{figure}[h]
\centering
\fbox{\includegraphics[width=10cm]{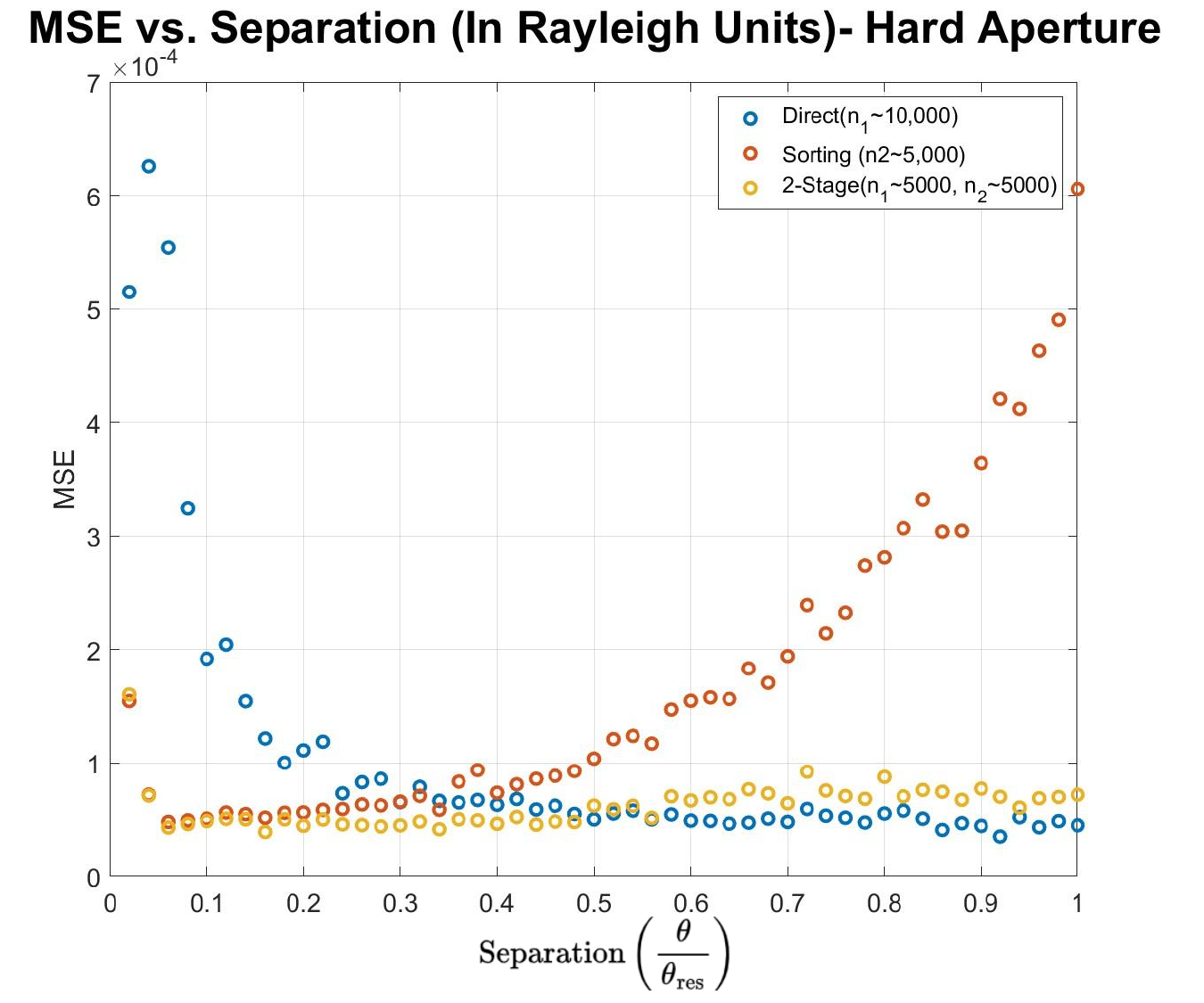}}
\caption{Monte-Carlo simulation that shows the mean squared error (MSE) of separation estimation using 10000 photons of direct imaging (blue circles) {(refer to section 5.2 for further details)}, 5000 photons in direct imaging for centroid estimation, and 5000 photons for sorting (orange circles) {(refer to section 5.3.1 for further details)}, and 5000 photons in direct imaging for centroid estimation and for separation estimation, and 5000 photons for sorting (yellow circles) {(refer to section 5.3.3 for further details)}.{Not shown on the graph, yet important to note is the quantum Cramer-Rao bound for MSE that is given as follows:$\frac{4\theta_{\rm res}^2}{N}=6.4904e-08$, where $\sigma$ is defined in equation \ref{eq:theta_res},and $N$ is the total number of photons. } Note that because in our simulation we assumed the absence of bias the plot for the MSE vs. the separation will look identical to the plot for variance vs. separation}
\label{fig:Monte Carlo}
\end{figure}

\begin{figure*}[!ht]
\centering
\fbox{\includegraphics[width=\linewidth]{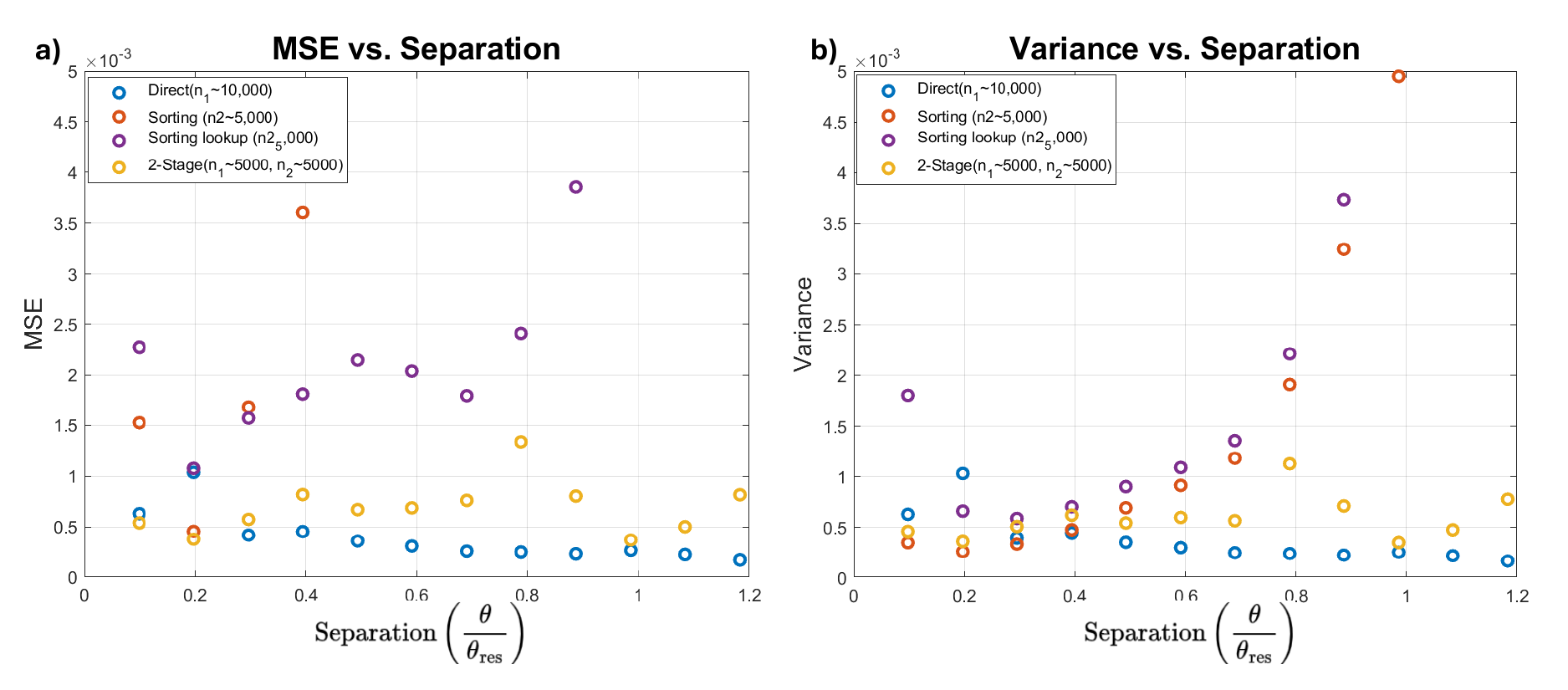}}
\caption{a) Mean squared error of the separation estimation , b) Variance of the separation estimation using 10000 photons of direct imaging (blue circles) {(refer to section~\ref{sec:Statistics} for further details)}, 5000 photons in direct imaging for centroid estimation and 5000 photons for sorting with the maximum likelihood method (orange circles) {(refer to section 5.3.1 for further details)}, 5000 photons in direct imaging for centroid estimation and 5000 photons for sorting with the comparison to characterization method (purple circles) {(refer to section 5.3.2 for further details)}, {and 5000 photons in direct imaging for centroid estimation and for separation estimation, 5000 photons for sorting with the maximum likelihood method (yellow circles)(refer to section 5.3.3 for further details).}
}
\label{fig:MSE_direct_sorting}
\end{figure*}
{Before conducting the experiment, we ran a Monte Carlo simulation, which replicates the results of~\cite{Grace:20} with the hard circular aperture, simulating the estimation procedure for direct imaging, mode sorting, and a two-stage receiver with equal time-share between the two.} The results of the simulation are promising as the separation estimation MSE in the deep sub-Rayleigh regime (below 0.5 Rayleigh units of separation) shows an improvement by a factor of up to 12.7. Note that in the Monte Carlo simulation, we assumed a perfect mode-sorter, with no crosstalk, (see supplementary information), and a perfect detector (no noise). {This simulation comes to show what MSE an ideal mode sorter can achieve}. The simulation results can be seen in Fig. \ref{fig:Monte Carlo}. 
In this figure, when looking at the simulation results when only using the sorting data for separation estimation (orange circles), one can notice a counter-intuitive phenomenon, the MSE increases with the separation. This cam be explained by the fact that at very small separations most of the Fisher information exists in the modes that we sort (PAD00 and PAD01) and as the separation increases there is more Fisher information that lives in higher order modes. Another way to think about it is by looking at fig.\ref{fig:Lookup_table}. When the separation is very small, there are very few photons arriving at the PAD01 spot and every change in that is easily detected. However, as the separation increases, small changes at the intensity of the PAD01 spot become harder to detect and therefore making an error becomes more likely.

{Figs.\ref{fig:MSE_direct_sorting}a and \ref{fig:MSE_direct_sorting}b present our experimental results for point source separation estimation in terms of the estimator MSE and variance, respectively. In a real-world imaging scenario, MSE is the most relevant quantity, as it directly quantifies the average error from the true value of the parameter. For 0.2 Rayleigh units of separation, our two-stage receiver achieves an MSE 2.76 times lower than experimental direct imaging.} 

{However, we do not see a consistent trend of sub-Rayleigh superiority over direct imaging in terms of MSE in Fig.~\ref{fig:MSE_direct_sorting}. To understand this, we can recall from Eq.~\eqref{eq:MSE_Var_Bias} that $\textrm{MSE}(\hat{\theta}_{\rm meas})=\textrm{Var}(\hat{\theta}_{\rm meas})+\textrm{B}(\hat{\theta}_{\rm meas})^2$. In Section 6 of the Supplementary Material, we show that our mode sorting measurement exhibited significant estimator bias. The weakness in accuracy of our second-stage measurement is a significant limiting factor in achieving advantageous MSE over direct imaging. This biased estimation is most likely a result of systematic errors in our experiment, e.g., unmitigated crosstalk between spatial modes.}

\begin{figure}[h]
\centering
\fbox{\includegraphics[width=12cm]{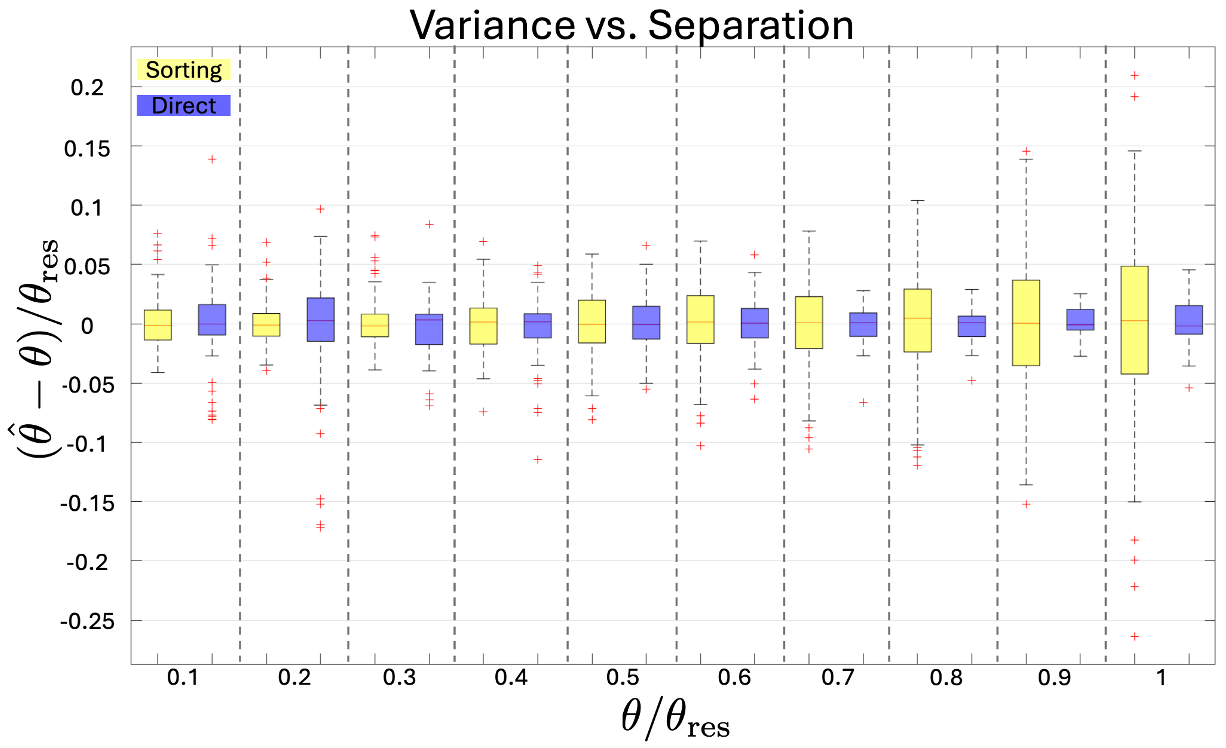}}
\caption{{This plot shows the variance of the separation estimation using sorting (yellow) and direct(blue). In this plot, we subtracted the mean of the of the estimation such that all bias is neglected in this plot. On each box, the central mark indicates the median, and the bottom and top edges of the box indicate the 25th and 75th percentiles, respectively. The whiskers extend to the most extreme data points not considered outliers, and the outliers are plotted individually using the '+' marker symbol.}}
\label{fig:Variance_comparison}
\end{figure}

{On the other hand, Fig.~\ref{fig:MSE_direct_sorting}b reveals that the estimator variance for our mode sorting measurement (which includes adaptive alignment from the stage-one centroid estimate) is consistently superior to direct imaging below 0.4 Rayleigh units of point source separation. The same results, neglecting the bias, are shown in Fig.~\ref{fig:Variance_comparison}. In that figure, one can see a more statistically descriptive plot of the separation estimation variance using direct imaging and mode sorting. In this plot, it becomes apparent that at the deep Rayleigh regime, our mode sorter will give more consistent results with a smaller spread of the estimates}. 

{Estimator variance is the statistic that quantifies the precision achievable with a particular measurement scheme, and the most common information theoretic measure for parameter estimation, the Fisher information, is used to bound the estimator variance of any unbiased estimator~\cite{VanTrees2010} In other words, while our experimental data may be corrupted by systematic errors that produce estimator bias, the superior estimator variance reveals that our experiment succeeds in demonstrating an adaptive measurement scheme with fundamentally reduced random noise compared with direct imaging for sub-diffraction parameter estimation. Under optimized experimental conditions we expect that our informationally advantageous measurement scheme would reliably produce reduced MSEs for sub-Rayleigh imaging.} Some possible sources of systematic error that may have impeded our estimation accuracy in this experiment include: crosstalk, although it was taken into account in the data processing, misalignment of the system, generally and with respect to the centroid, and shot noise in the detector (see Supplementary Material Section 6 for further discussion of the sources of error).

{When comparing Fig.\ref{fig:MSE_direct_sorting}a and Fig.\ref{fig:MSE_direct_sorting}b it is notable that the characterization method (purple) performs better than sorting (orange) in MSE, but worse in variance. We believe that it performs better in MSE since the \emph{in situ} characterization inherently corrects for systematic errors leading to estimator bias (see Supplementary Materials Section 6). However, the variance is larger in this method due to the discretization of the possible separations that this estimator can choose, reducing estimation precision.} 

\begin{figure}[h]
\centering
\fbox{\includegraphics[width=10cm]{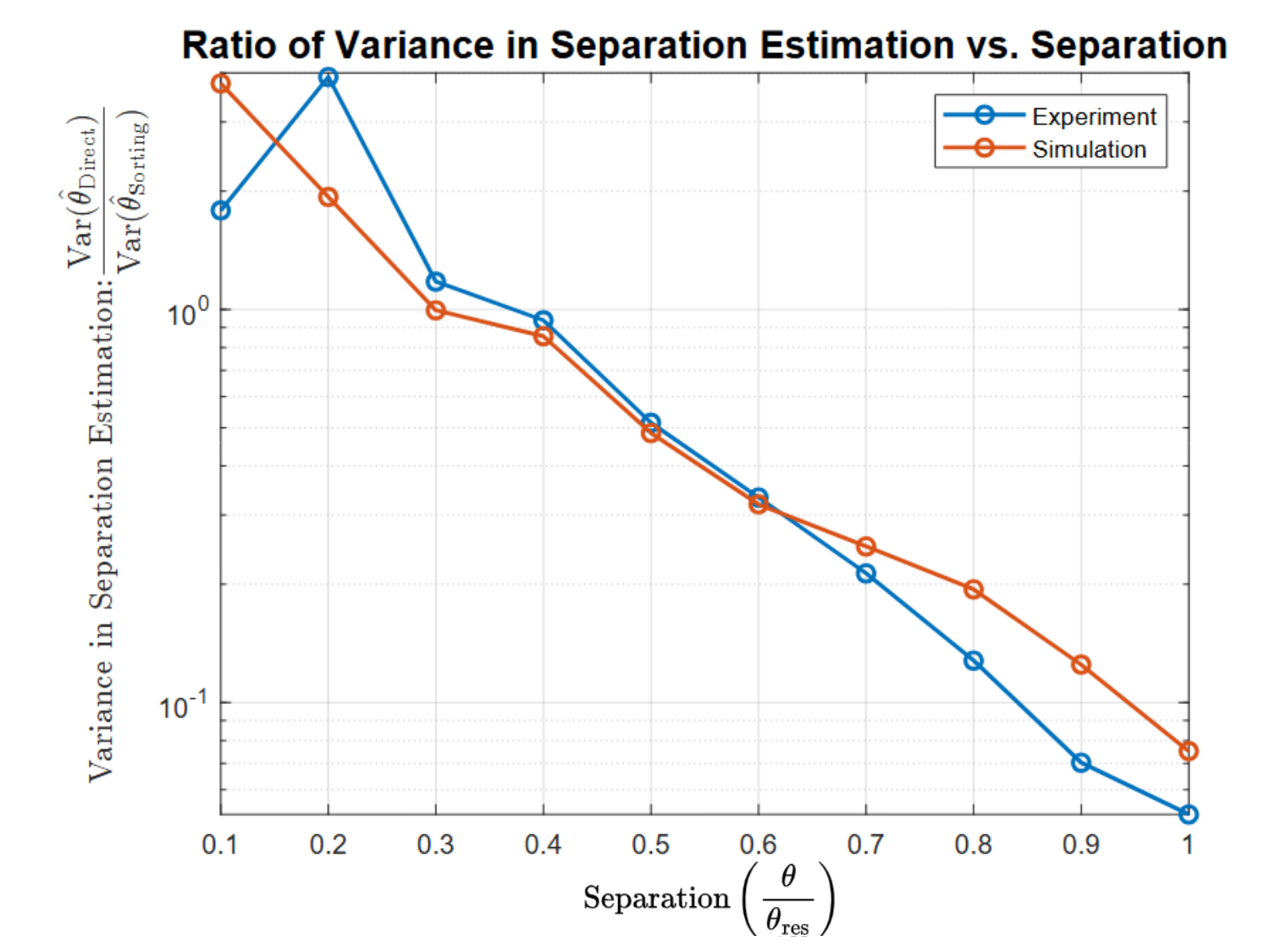}}
\caption{{Ratio of the variance of separation estimation with direct imaging over the variance of separation estimation with mode sorting as a function of separation in our experiment (blue) and in our Monte-Carlo simulation (orange).} }
\label{fig:Ratio_Variance}
\end{figure}

{As a final evaluation of the estimation precision achievable with our adaptive receiver, Fig.~\ref{fig:Ratio_Variance} directly compares the variance of our experimental estimator against that of our Monte Carlo simulation. Specifically, Fig.~\ref{fig:Ratio_Variance} plots the ratios between the estimator variances of direct imaging and two-stage mode sorting and compares that ratio for our experiment and simulation. This visualization of our data reveals that, while the idealized simulation understandably produces lower errors than our experiment, the factor by which the mode sorting receiver over- or underperforms direct imaging is largely consistent between simulation and experiment. Fig.~\ref{fig:Ratio_Variance} shows that our experiment demonstrated roughly all of the relative gain in estimation precision over direct imaging that should be expected based on the simulation results.}

\section{Conclusion}
{Multi-plane light conversion (MPLC) involves passing a multi-spatial mode optical field through multiple stages of programmed spatial-phase-modulation interspersed with fixed-length propagation segments in an isotropic medium. With sufficiently many stages, and sufficiently large transverse cross sections of spatial phase modulation (so as to avoid truncation of the propagating field), MPLC allows for the realization of a general unitary transformation on the span of a set of co-propagating orthogonal spatial modes. We built an MPLC where the spatial phase modulation of the field was realized using distinct segments of a single spatial light modulator (SLM). Using this MPLC, we realized a fully-reconfigurable spatial mode sorter, i.e., a device that spatially separates a number of mutually orthogonal spatial mode functions of any specified modal basis, into spatially-separated beams which can then be detected using individual detectots or assigned detector-pixel segments of a pixelated camera. We showed sorting various orthogonal mode sets, including the binary-phase Hadamard mode sets, the 2D Hermite-Gauss modes, and what we call the PSF-adapted (PAD) basis modes corresponding to the hard-circular-aperture pupil of our imaging system. We showed that our reconfigurable mode sorter can sort two orthogonal modes with less than $1\%$ power cross-talk, and can sort four modes with about $5\%$ power cross-talk. Using this mode sorter, we built a partially-emulated experiment to demonstrate improved spatial resolution---i.e., lower-variance estimation of the angular separation between two equally-bright incoherently-radiating quasi-monochromatic point sources with a sub-Rayleigh separation---over using direct detection using a conventional image-plane focal-plane array. Our receiver follows the recipe from Ref.~\cite{Grace:20} where the receiver starts out with direct detection as the first stage to estimate the centroid of the two-point-source constellation, and switches to a three-element PAD-basis MPLC mode sorter---at half the total allocated integration time---whose center was electronically-programmed, leveraging the programmability of the MPLC, using the (noisy) estimate of the centroid obtained from the first stage. While our experiment made limited use of the reconfigurability of our MPLC mode sorter, the ability to switch between modal bases is applicable in more challenging multi-source localization tasks with more unknown prior information in subsequent investigations~\cite{Kwan23}. Building an MPLC-assisted super-resolution imaging system that outperforms direct imaging despite the additional losses internal to the MPLC remains an important future direction. Finally, our experiment used a digital-domain incoherent combination of the detection outcomes---both for direct detection as well as for mode sorting---from a single spatially-scanned laser-light source to mimic the intensity pattern. Extending this work to an experiment with a thermal light and a pinhole mask to mimic a true multi-point source scene, remains an important subject of future research.}

\FloatBarrier
\section*{Funding}
DARPA IAMBIC Program under Contract No. HR00112090128.
\section*{Acknowledgments}
This research was supported by the DARPA IAMBIC Program under Contract No. HR00112090128. The
views, opinions and/or findings expressed are those of the authors and should not be interpreted as representing the official views or policies of
the Department of Defense or the U.S. Government.
\section*{Disclosures}
The authors declare no conflicts of interest.

\section*{Data availability}
Data underlying the results presented in this paper are not publicly available at this time due to the large amount and size of the raw data, but may be obtained from the authors upon reasonable request.\\

See Supplement 1 for supporting content

\bibliography{Bibliography}

\begin{thebibliography}{10}
\newcommand{\enquote}[1]{``#1''}

\bibitem{Grace:20}
M.~R. Grace, Z.~Dutton, A.~Ashok, and S.~Guha, \enquote{Approaching quantum-limited imaging resolution without prior knowledge of the object location,} {\protect\JournalTitle{J. Opt. Soc. Am. A}} \textbf{37}, 1288--1299 (2020).

\bibitem{barentine2023integrated}
A.~E. Barentine, Y.~Lin, E.~M. Courvan, \emph{et~al.}, \enquote{An integrated platform for high-throughput nanoscopy,} {\protect\JournalTitle{Nature Biotechnology}} \textbf{41}, 1549--1556 (2023).

\bibitem{hampson2021adaptive}
K.~M. Hampson, R.~Turcotte, D.~T. Miller, \emph{et~al.}, \enquote{Adaptive optics for high-resolution imaging,} {\protect\JournalTitle{Nature Reviews Methods Primers}} \textbf{1}, 68 (2021).

\bibitem{VanTrees2010}
H.~L. {Van Trees} and K.~L. Bell, \enquote{{Lower Bounds for Parametric Estimation with Constraints},} {\protect\JournalTitle{Bayesian Bounds for Parameter Estimation and Nonlinear Filtering/Tracking}}  (2010).

\bibitem{berger1999integrated}
J.~O. Berger, B.~Liseo, and R.~L. Wolpert, \enquote{Integrated likelihood methods for eliminating nuisance parameters,} {\protect\JournalTitle{Statistical science}} \textbf{14}, 1--28 (1999).

\bibitem{Rayleigh}
Rayleigh, \enquote{Xxxi. investigations in optics, with special reference to the spectroscope,} {\protect\JournalTitle{The London, Edinburgh, and Dublin Philosophical Magazine and Journal of Science}} \textbf{8}, 261--274 (1879).

\bibitem{harootunian1986super}
A.~Harootunian, E.~Betzig, M.~Isaacson, and A.~Lewis, \enquote{Super-resolution fluorescence near-field scanning optical microscopy,} {\protect\JournalTitle{Applied Physics Letters}} \textbf{49}, 674--676 (1986).

\bibitem{hell1994breaking}
S.~W. Hell and J.~Wichmann, \enquote{Breaking the diffraction resolution limit by stimulated emission: stimulated-emission-depletion fluorescence microscopy,} {\protect\JournalTitle{Optics letters}} \textbf{19}, 780--782 (1994).

\bibitem{betzig2006imaging}
E.~Betzig, G.~H. Patterson, R.~Sougrat, \emph{et~al.}, \enquote{Imaging intracellular fluorescent proteins at nanometer resolution,} {\protect\JournalTitle{science}} \textbf{313}, 1642--1645 (2006).

\bibitem{rust2006sub}
M.~J. Rust, M.~Bates, and X.~Zhuang, \enquote{Sub-diffraction-limit imaging by stochastic optical reconstruction microscopy (storm),} {\protect\JournalTitle{Nature methods}} \textbf{3}, 793--796 (2006).

\bibitem{hess2006ultra}
S.~T. Hess, T.~P. Girirajan, and M.~D. Mason, \enquote{Ultra-high resolution imaging by fluorescence photoactivation localization microscopy,} {\protect\JournalTitle{Biophysical journal}} \textbf{91}, 4258--4272 (2006).

\bibitem{gustafsson2000surpassing}
M.~G. Gustafsson, \enquote{Surpassing the lateral resolution limit by a factor of two using structured illumination microscopy,} {\protect\JournalTitle{Journal of microscopy}} \textbf{198}, 82--87 (2000).

\bibitem{19}
M.~Tsang, R.~Nair, and X.-M. Lu, \enquote{Quantum theory of superresolution for two incoherent optical point sources,} {\protect\JournalTitle{Phys. Rev. X}} \textbf{6}, 031033 (2016).

\bibitem{Nair:16}
R.~Nair and M.~Tsang, \enquote{Interferometric superlocalization of two incoherent optical point sources,} {\protect\JournalTitle{Opt. Express}} \textbf{24}, 3684--3701 (2016).

\bibitem{Paur:16}
M.~Pa\'{u}r, B.~Stoklasa, Z.~Hradil, \emph{et~al.}, \enquote{Achieving the ultimate optical resolution,} {\protect\JournalTitle{Optica}} \textbf{3}, 1144--1147 (2016).

\bibitem{Tan:23}
X.-J. Tan, L.~Qi, L.~Chen, \emph{et~al.}, \enquote{Quantum-inspired superresolution for incoherent imaging,} {\protect\JournalTitle{Optica}} \textbf{10}, 1189--1194 (2023).

\bibitem{Rouviere:24}
C.~Rouvi\`{e}re, D.~Barral, A.~Grateau, \emph{et~al.}, \enquote{Ultra-sensitive separation estimation of optical sources,} {\protect\JournalTitle{Optica}} \textbf{11}, 166--170 (2024).

\bibitem{Santamaria:23}
L.~Santamaria, D.~Pallotti, M.~S. de~Cumis, \emph{et~al.}, \enquote{Spatial-mode demultiplexing for enhanced intensity and distance measurement,} {\protect\JournalTitle{Opt. Express}} \textbf{31}, 33930--33944 (2023).

\bibitem{Boucher:20}
P.~Boucher, C.~Fabre, G.~Labroille, and N.~Treps, \enquote{Spatial optical mode demultiplexing as a practical tool for optimal transverse distance estimation,} {\protect\JournalTitle{Optica}} \textbf{7}, 1621--1626 (2020).

\bibitem{Fontaine_Ryf_Chen_Neilson_Kim_Carpenter_2019}
N.~K. Fontaine, R.~Ryf, H.~Chen, \emph{et~al.}, \enquote{Laguerre-gaussian mode sorter,} {\protect\JournalTitle{Nature Communications}} \textbf{10} (2019).

\bibitem{morizur2010programmable}
J.-F. Morizur, L.~Nicholls, P.~Jian, \emph{et~al.}, \enquote{Programmable unitary spatial mode manipulation,} {\protect\JournalTitle{JOSA A}} \textbf{27}, 2524--2531 (2010).

\bibitem{labroille2014efficient}
G.~Labroille, B.~Denolle, P.~Jian, \emph{et~al.}, \enquote{Efficient and mode selective spatial mode multiplexer based on multi-plane light conversion,} {\protect\JournalTitle{Optics express}} \textbf{22}, 15599--15607 (2014).

\bibitem{Grenapin23}
F.~Grenapin, D.~Paneru, A.~D'Errico, \emph{et~al.}, \enquote{Superresolution enhancement in biphoton spatial-mode demultiplexing,} {\protect\JournalTitle{Phys. Rev. Appl.}} \textbf{20}, 024077 (2023).

\bibitem{de2021discrimination}
J.~De~Almeida, J.~Ko{\l}ody{\'n}ski, C.~Hirche, \emph{et~al.}, \enquote{Discrimination and estimation of incoherent sources under misalignment,} {\protect\JournalTitle{Physical Review A}} \textbf{103}, 022406 (2021).

\bibitem{bao2021quantum}
F.~Bao, H.~Choi, V.~Aggarwal, and Z.~Jacob, \enquote{Quantum-accelerated imaging of n stars,} {\protect\JournalTitle{Optics Letters}} \textbf{46}, 3045--3048 (2021).

\bibitem{Kwan23}
K.~K. Lee, C.~N. Gagatsos, S.~Guha, and A.~Ashok, \enquote{Quantum-inspired multi-parameter adaptive bayesian estimation for sensing and imaging,} {\protect\JournalTitle{IEEE Journal of Selected Topics in Signal Processing}} \textbf{17}, 491--501 (2023).

\bibitem{fickler2017custom}
R.~Fickler, M.~Ginoya, and R.~W. Boyd, \enquote{Custom-tailored spatial mode sorting by controlled random scattering,} {\protect\JournalTitle{Physical Review B}} \textbf{95}, 161108 (2017).

\bibitem{Matlin2022}
E.~F. Matlin and L.~J. Zipp, \enquote{Imaging arbitrary incoherent source distributions with near quantum-limited resolution,} {\protect\JournalTitle{Scientific Reports}} \textbf{12}, 2810 (2022).

\bibitem{martin2018photonic}
A.~Martin, D.~Dodane, L.~Leviandier, \emph{et~al.}, \enquote{Photonic integrated circuit-based fmcw coherent lidar,} {\protect\JournalTitle{Journal of Lightwave Technology}} \textbf{36}, 4640--4645 (2018).

\bibitem{fang2024nonvolatile}
Z.~Fang, R.~Chen, J.~E. Froch, \emph{et~al.}, \enquote{Nonvolatile phase-only transmissive spatial light modulator with electrical addressability of individual pixels,} {\protect\JournalTitle{ACS nano}} \textbf{18}, 11245--11256 (2024).

\bibitem{Aqil}
A.~Sajjad, M.~R. Grace, Q.~Zhuang, and S.~Guha, \enquote{Attaining quantum limited precision of localizing an object in passive imaging,} {\protect\JournalTitle{Phys. Rev. A}} \textbf{104}, 022410 (2021).

\bibitem{Ozer:thesis}
I.~Ozer, \enquote{Reconfigurable mode sorter for super resolution imaging,} Master's thesis, The University of Arizona, Tucson, AZ (2022). 2022.29325286.

\bibitem{Kerviche2017a}
R.~Kerviche, S.~Guha, and A.~Ashok, \enquote{{Fundamental limit of resolving two point sources limited by an arbitrary point spread function},} {\protect\JournalTitle{IEEE International Symposium on Information Theory - Proceedings}}  (2017).

\bibitem{Rehacek2017b}
J.~Rehacek, M.~Paúr, B.~Stoklasa, \emph{et~al.}, \enquote{Optimal measurements for resolution beyond the rayleigh limit,} {\protect\JournalTitle{Opt. Lett.}} \textbf{42}, 231–234 (2017).

\bibitem{YuPrasad2019}
Z.~Yu and S.~Prasad, \enquote{Quantum limited superresolution of an incoherent source pair in three dimensions,} {\protect\JournalTitle{Phys. Rev. Lett.}} \textbf{121}, 180504 (2018).

\end{thebibliography}






\end{document}